\documentclass[10pt,a4paper,5p,sort&compress]{elsarticle} 
\usepackage{amsfonts}
\usepackage{textcomp}
\usepackage{amsmath}
\usepackage{bm}
\usepackage{xcolor}
\usepackage[hyperindex,breaklinks]{hyperref}
\hypersetup{
    colorlinks,
    linkcolor={red!50!black},
    citecolor={blue!50!black},
    urlcolor={blue!80!black}
}
\usepackage{graphicx}
\usepackage{url}
\usepackage[utf8]{inputenc}
\usepackage{comment}
\usepackage{lineno}
\usepackage{orcidlink}

\usepackage{enumitem}

\journal{Nuclear Instruments and Methods in Physics Research Section A}

\usepackage[singlelinecheck=false]{caption}
\begin{document}
\begin{frontmatter}
\title {
Seven-dimensional Trajectory Reconstruction for VAMOS++ 
}

\author{M.~Rejmund\,\orcidlink{0009-0009-8626-8756}}
\author{A.~Lemasson\,\orcidlink{0000-0002-9434-8520}} 

\address{GANIL, CEA/DRF - CNRS/IN2P3, Bd Henri Becquerel, BP 55027,
  F-14076 Caen Cedex 5, France}
\begin{abstract}

The VAMOS++ magnetic spectrometer is characterized by a large angular and momentum acceptance and highly 
non-linear ion optics properties requiring the use of software ion trajectory reconstruction 
methods to measure the ion magnetic rigidity and the trajectory length between the beam interaction point and 
the focal plane of the spectrometer. 
Standard measurements, involving the use of a thin target and a narrow beam spot, allow the assumption of a point-like beam interaction 
volume for ion trajectory reconstruction. However, this represents a limitation for the case of large beam spot size or extended gaseous 
target volume. To overcome this restriction, a seven-dimensional reconstruction method incorporating the reaction position coordinates 
was developed, making use of artificial deep neural networks. The neural networks were trained on a theoretical dataset generated by 
standard magnetic ray-tracing code. Future application to a voluminous gas target, necessitating the explicit inclusion of 
the three-dimensional position of the beam interaction point within the target in the trajectory reconstruction method, is discussed.
The performances of the new method are presented along with a comparison of mass resolution obtained with previously reported model 
for the case of thin-target experimental data. 


\end{abstract}
\end{frontmatter}

\section*{Introduction}

The large angular and momentum acceptance  spectrometer VAMOS++~\cite{Pullanhiotan2008, Rejmund2011} is widely 
utilized in research at beam energies near the Coulomb barrier, particularly in domains such as nuclear structure and 
reaction mechanism. Its primary function is to provide the isotopic identification of incoming ions of interest, specifically 
their atomic mass number $A$ and atomic number $Z$.

The key observable provided by the magnetic spectrometer is the magnetic rigidity $B\rho$ of the ion, which is determined 
based on the dispersive action of the magnetic dipole. The atomic mass number $A$ can then be calculated by combining 
the magnetic rigidity $B\rho$ with measurements of the ion’s velocity $v$, the trajectory length between the beam interaction 
point and the focal plane of the spectrometer $l$, and the total energy $E_{tot}$ for the detected ions. 
For further details, refer to Ref.~\cite{Lemasson2023a}. The atomic number $Z$ can be obtained by correlating the energy 
loss $\Delta E$ with the total energy $E_{tot}$ and the atomic charge state $q$ by combining the total energy $E_{tot}$,
$B\rho$  and $v$.

The large angular and momentum acceptance of the VAMOS++ spectrometer is associated with highly nonlinear ion optics. 
The measurement of the magnetic rigidity $B\rho$ and trajectory length between the beam interaction point and the focal plane 
of the spectrometer $l$ requires the complex measurements of the initial (at the entrance) and final (in the focal plane of 
the spectrometer) coordinates and the trajectory reconstruction algorithms. 
The initial coordinates are determined using the dual position-sensitive Multi-Wire Proportional Chamber (MWPC) telescope~\cite{Vandebrouck2016}, situated at the entrance of VAMOS++. The final coordinates are measured by two 
large-area position-sensitive MWPCs positioned within the focal plane. 
Each MWPC provides horizontal and vertical position as well as timing.
Several trajectory reconstruction algorithms have been utilized at VAMOS++~\cite{Lemasson2023a}, of which the most advanced 
and performant is the four-dimensional 4D mapping method, based on the large dimension four-dimensional 
arrays indexed by 
the initial and final coordinates. 
However, this method assumes a point-like beam spot. 
This presents a limitation in the case of large beam spot size or extended gaseous target volume.
One of the upcoming experimental programs at VAMOS++ aims to use a voluminous gas target 
designed to study fission dynamics. 
In this case the optical axis of the spectrometer will be rotated relative to the beam axis.
Two fission products will be detected in coincidence, one in VAMOS++ and another in the detector analogous to  
the dual position-sensitive MWPC telescope~\cite{Vandebrouck2016} positioned at the entrance of VAMOS++. 
The combination of data obtained from two telescopes and the use of two-body kinematics
will enable the determination of the three-dimensional position of the beam interaction point within the target.
In this context, it is necessary to incorporate the beam interaction coordinates into the trajectory reconstruction algorithm. 
The direct extension of the 4D method to account for this is not feasible  due to the excessively large dimensions of 
the matrices, as inferred from Ref.~\cite{Lemasson2023a}.
The goal of this work was to develop a novel trajectory reconstruction method capable of handling the seven-dimensional 
highly nonlinear problem. This was achieved by employing artificial deep neural networks, which were trained on the theoretical 
dataset of ion trajectories calculated by the standard magnetic ray-tracing code.

\begin{figure}[t]\center
\includegraphics[width=1\columnwidth]{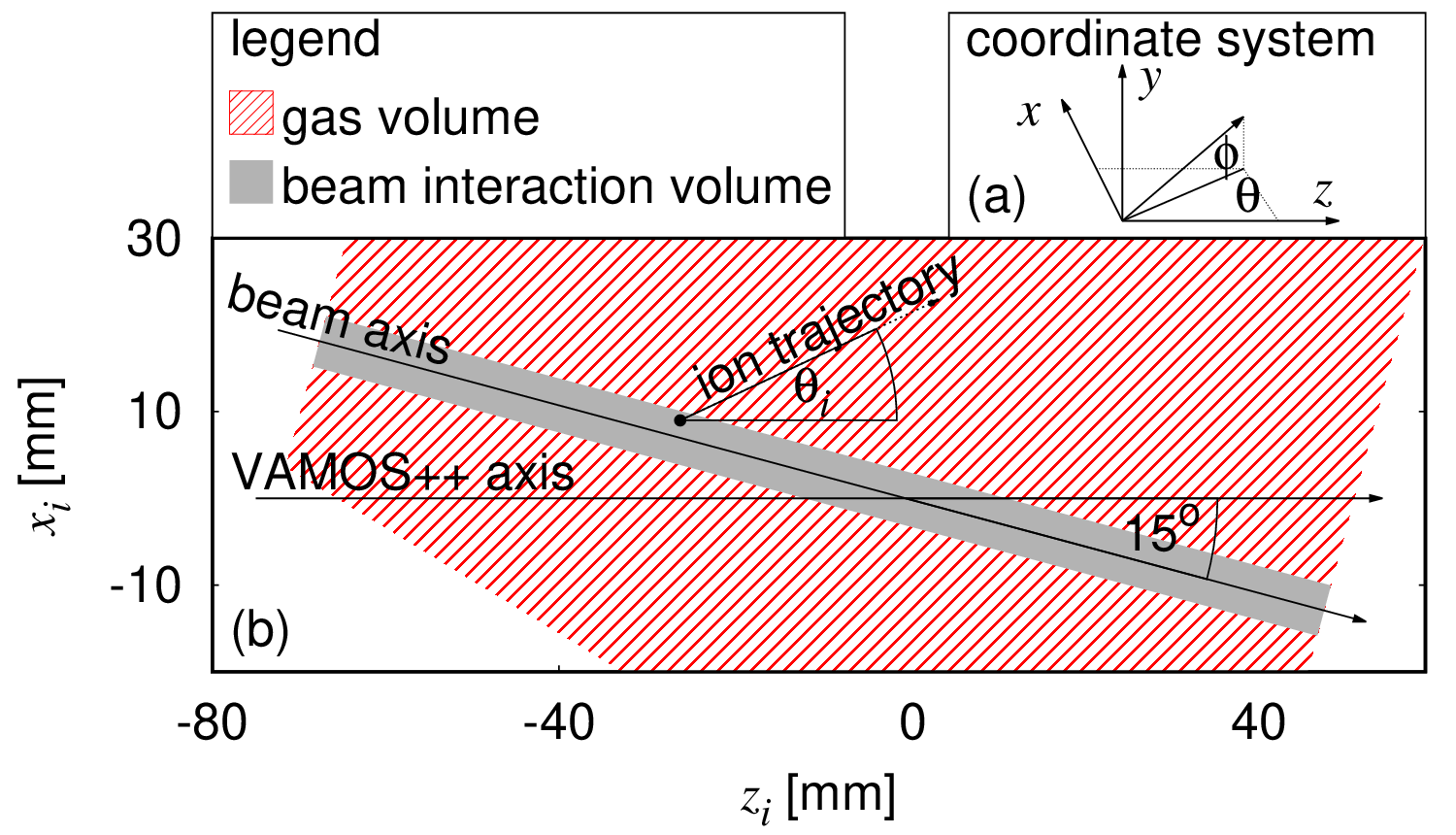}
\caption{
Beam interaction in the gas volume:
(a) Standard coordinate system: The coordinate system is defined with the $xz$ plane coinciding with the dispersive plane of 
the spectrometer.
(b) Sectional view in the $xz$ plane at the VAMOS++ entrance: A sectional view of the $xz$ plane at the entrance of VAMOS++ 
is presented. The origin of the coordinate system corresponds to the intersection of the beam axis and VAMOS++ optical axis.
The angle between the optical axis and the beam axis is $15^\circ$. The total gas volume and the beam-gas 
interaction volume are indicated by hatched and grey areas, respectively. An illustration of an outgoing ion originating from 
the beam-gas interaction is also provided.
\label{fig:GasCell}}
\end{figure}

\begin{figure}[t]
\includegraphics[width=1\columnwidth]{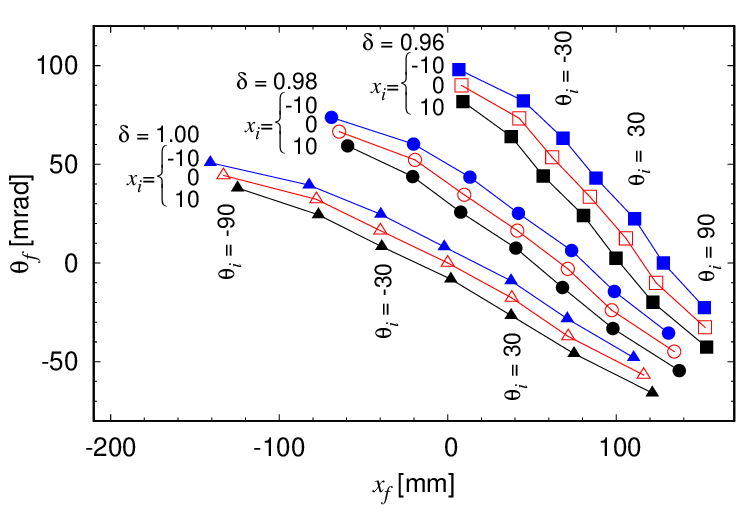}
\caption{
Aberrations of the VAMOS++ spectrometer with variable horizontal interaction position:
The calculated angle, $\theta_{f}$, as a function of the position, $x_{f}$, at the image focal plane of VAMOS++ demonstrates 
the effects of aberrations in both position and angle. The figure presents the final trajectories for varying relative rigidity, 
$\delta = B\rho/B\rho_0$, set to  $0.94$, $0.98$ and $1.00$, horizontal angle, $\theta_{i}$, spanning from $-90$ to $90$~mrad 
in increments of 30 mrad, horizontal 
position, $x_{i}$, set to -10, 0, and 10 mm, vertical and longitudinal positions, $y_{i}$ and $z_{i}$, both set to $0$~mm, and 
vertical angle, $\phi_{i}$, set to $0$~mrad.
\label{fig:TfXf}}
\end{figure}

\section*{Experiments with extended beam interaction volume}

The details of the standard coordinate system employed with VAMOS++ are presented in Fig.~\ref{fig:GasCell}(a).
The horizontal $xz$ plane corresponds to the dispersive plane of the spectrometer. At the entrance of the spectrometer, 
in the VAMOS++ coordinate system, the $z$ axis is parallel to the optical axis of the spectrometer and is designated by 
the indexes $i$. The VAMOS++ focal plane coordinate system is typically rotated about the $y$ axis by $45^\circ$ with 
respect to the coordinates at the entrance and translated to the focal plane following the deviation of the nominal trajectory. 
It is denoted by the indexes $f$. The coordinates indexed with $b$ refer to the beam coordinate system, in which the $z$ axis 
is parallel to the beam axis.

The experimental setup is depicted in Fig.~\ref{fig:GasCell}(b), which presents a sectional view of the gas target volume 
in the horizontal $xz$ plane. The angle between the optical axis and the beam axis is $15^\circ$. For experiments involving 
a thin target, the beam spot is positioned at $(x_i, y_i, z_i) = (0, 0, 0)$ and typically has a diameter of $\sigma(x_i) = 0.50$~mm 
and $\sigma(y_i) = 0.65$~mm~\cite{Vandebrouck2016}.
The hatched area represents the total gas volume, while the gray area denotes the volume considered for the beam-gas interaction. 
To account for beam profile broadening or displacement, the considered beam-gas interaction volume is defined in beam coordinates 
as follows: $x_b : [-3, 3]$~mm, $y_b: [-6, 6]$~mm, and $z_b : [-70, 50]$~mm.
While variations in the $z_i$ direction primarily affect the trajectory length $l_f$ between the interaction point and the focal plane 
of the spectrometer, significant variations in the $x_i$ direction substantially influence the focal plane image obtained.

\section*{Optical aberrations}

The overview of the focal plane aberrations of VAMOS++, assuming point-like beams ($\sigma(x_i)=\sigma(y_i)=\sigma(z_i)=0$), can be found in 
Fig.~2 of Ref.~\cite{Lemasson2023a}.
We will focus on the focal plane image related to a variation in the dispersive horizontal $xz$ plane of the interaction position, $x_i$.
As depicted in Fig.~\ref{fig:GasCell}(b), the coordinate $x_i$ can range from approximately $-15$~mm to $20$~mm, 
while the coordinate $z_i$ can range from approximately $-70$~mm to $50$~mm.
Fig.~\ref{fig:TfXf} presents the focal plane image for three relative rigidities defined as $\delta = B\rho/B\rho_0$, 
$\delta = 0.94, 0.98, 1.00$, three horizontal positions, 
$x_{i}=-10,0,10$~mm, and the horizontal angle $\theta_{i}$ in the range from $-90$~mrad to $90$~mrad, with increments of $30$~mrad. 
These images were calculated using the ray-tracing code ZGOUBI~\cite{Meot1999}.
It can be seen that a change in the $x_i$ coordinate by only $10$~mm results in a significant displacement compared to a change in 
the relative rigidity $\delta$ by $2\%$, as presented. This highlights the necessity of explicitly incorporating the three-dimensional 
position of the beam interaction point within the target into the trajectory reconstruction method.
Note, that VAMOS++ typically operates in the sub-percent resolution regime in terms of the full width at half maximum 
FWHM$(A)/A$~\cite{Lemasson2023a}.

\section*{Deep neural network model of the ion optics}
\subsection*{General network architecture}
\begin{figure}[t]
\includegraphics[width=1\columnwidth]{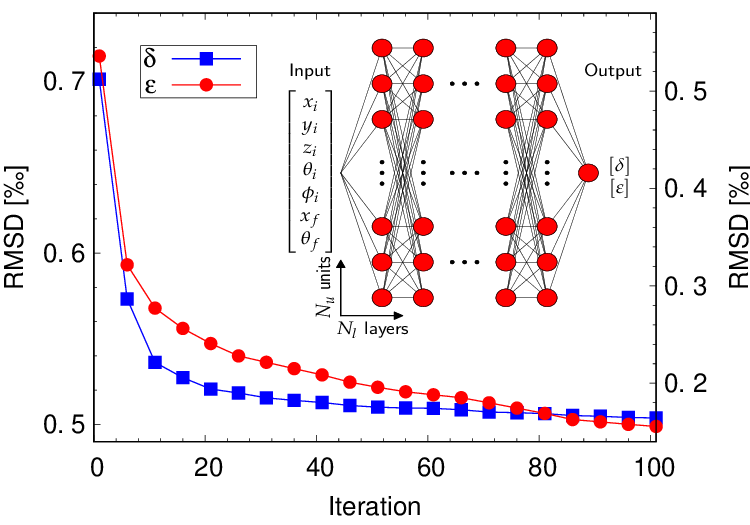}
\caption{Training convergence:
The root mean square deviation (RMSD) for $\delta$ (blue squares, left axis) and $\varepsilon$ (red circles, right axis) is plotted 
as a function of the number of iterations for the neural network architecture $N_l \times N_u = 6 \times 32$.
Inset: Deep neural network architecture: $N_l \times N_u$, composed of $N_l$ layers and $N_u$ units (neurons) per layer, 
followed by a single output unit.
The deep neural network input and output variables are also indicated (see text).
\label{fig:Conv}}
\end{figure}

The selection of artificial deep neural networks for trajectory reconstruction appears to be an optimal choice.
These networks emulate biological brain neurons and possess the ability to discern complex patterns.
We have opted for a dense deep feed-forward neural network architecture~\cite{kinsley2020,Subasi2020} of the form $N_l \times N_u$, comprising $N_l$ 
layers and $N_u$ units (neurons) per layer, followed by a single output unit. The schematic representation of the $N_l \times N_u$ 
Deep Neural Network (DNN) is depicted in the inset of Fig.~\ref{fig:Conv}.
The input to the DNN encompasses the following ion trajectory coordinates:
the interaction position of the beam within the target volume ($x_{i}, y_{i}, z_{i}$);
the horizontal and vertical angles of the outgoing reaction products ($\theta_{i}, \phi_{i}$);
the horizontal position and angle of the products at the focal plane of the spectrometer ($x_{f}, \theta_{f}$).
The output of the DNN provides the relative magnetic rigidity $\delta = B\rho/B\rho_0$, with the typical 
$B\rho_0 \sim 1$~Tm, 
or relative trajectory length between the interaction point 
and the focal plane of the spectrometer $\varepsilon = l_f/l_0$, where $l_0 = 760$~cm.
The DNN models employed for $\delta$ and $\varepsilon$ have the same architecture but are maintained separate for parallel execution. 
This approach will be referred to as the 7DNN trajectory reconstruction method.

\subsection*{Theoretical training dataset}

To train the DNN, the trajectory dataset was calculated using the ray-tracing code ZGOUBI~\cite{Meot1999}.
In total, we utilized $2\times 10^8$ trajectories that were randomly distributed over initial coordinates within the specified ranges: 
$\delta : [0.7,1.4]$, $\theta_i : [-150,150]$~mrad, $\phi_i : [-260,260]$~mrad and
$x_i, y_i$ and $z_i$ covering the beam-gas interaction volume as defined above in beam coordinates, 
$x_b : [-3,3]$~mm, $y_b: [-6,6]$~mm and $z_b : [-70,50]$~mm. 
It is noteworthy that $\delta_i = \delta_f = \delta$ and the index has been omitted.
The ray-tracing code ZGOUBI provided for each trajectory the corresponding final coordinates, $x_f$, $\theta_f$ and $\varepsilon$.
This dataset was designed to densely encompass the entire beam interaction volume, enabling the network 
to precisely map the diverse dependencies. In the event that interactions occur outside this volume or the VAMOS++ 
angle changes, the dataset must be adjusted, and the training process must be repeated.
The dataset of trajectories utilized is available at~\cite{Lemasson2025}. 

\subsection*{Training workflow}

The following workflow was used in this work:
\begin{itemize}
\item Network complexity evaluation:
Deep neural networks of varying complexities were evaluated, starting with the most complex network 
$N_l \times N_u = 12 \times 512$ and progressing to the least complex network of size $N_l \times N_u = 4 \times 16$. 
This progression involved decreasing $N_l$ by 2 and $N_u$ by a factor of 2 at each step. 
The objective was to select the lowest complexity network providing the best resolution in terms the FWHM$(A)/A$ 
when applied to the experimental validation dataset (see subsection~\nameref*{ExpVal}). 

\item Network training:
\begin{itemize}
\item The theoretical training dataset was randomly partitioned into the training set ($90\%$) and the validation set ($10\%$).

\item For every iteration: 
One iteration corresponds to the training of the DNN on the entire dataset for a single cycle.
The root mean square deviation (RMSD) for $\delta$ or $\varepsilon$ was monitored. 
In the present case, the RMSD for 
the training and validation sets decreased steeply with the number of iterations at the beginning of the training. 
After the extensive training period, the decrease in the RMSD for both sets became negligible. 
The case of overfitting, characterized by a decreasing RMSD for the training set and an increasing RMSD for the validation set, was not observed.

\item  Every 5 iterations: 
The network was applied to the experimental validation dataset. The resolution in terms of FWHM(A)/A was monitored 
for the ions in the range $A: [80:160]$ as a function of the position $x_f$ in steps of $20$~cm in the dispersive focal plane.
It has been observed that the networks attained their optimal performance for the high-rigidity side of the dispersive focal plane 
more rapidly than for the low-rigidity side. The achievable experimental resolution is limited by detector 
resolutions. 
The training was terminated when the  FWHM($A$)/$A$ no longer showed any further improvement,
as the subsequent convergence towards higher precision cannot be experimentally verified.

\end{itemize}
\end{itemize}

\subsection*{Experimental validation dataset}
\label{ExpVal}

To validate the 7DNN reconstruction capabilities we used the thin-target experimental data of 
the E826 GANIL experiment~\cite{DataE826}. In this experiment  the fission fragments were produced 
in fusion-fission and transfer-fission reaction-induced by  $^{238}$U beam at an energy of  $5.88$~MeV/u 
on the $0.5$~mg/cm$^2$ thick $^9$Be target. The VAMOS++ spectrometer used to detect and identify 
the fission fragments was positioned at $20^\circ$ relative 
to the beam axis. The 7DNN trained on the theoretical dataset along with the
the experimentally measured ($x_{i}$, $y_{i}$, $\theta_{i}$, $\phi_{i}$, $x_{f}$, $\theta_{f}$) coordinates was used
and $z_i = 0$ was assumed.
Typically $1\times 10^7$ events were used to evaluate the quality of the reconstruction in terms of  the full width at half maximum 
FWHM$(A)/A$.
The atomic mass number is obtained from the following relationships:
\begin{align}
\label{eq:eq1}
(A/q) &=  \frac{B\rho}{3.107 \cdot \beta \cdot \gamma} \nonumber\\
q_{int} &= \left\lfloor \frac{E_{tot}}{1~\text{u} \cdot (\gamma-1) \cdot (A/q)} +0.5\right\rfloor , \nonumber\\
A &= (A/q) \cdot q_{int}
\end{align}
where: $(A/q)$ is the mass-over-charge ratio, $q_{int}$ is the atomic charge state rounded up to nearest integer,  
$\text{u} = 931.494$~MeV/$c^2$ is the unified atomic unit, $\beta = v/c$ and
$\gamma$ is the Lorentz factor. With well-separated atomic charge states, as can be seen in Ref.~\cite{Kim2017}, 
the resolution of $A$ is primarily determined by the resolution of $B\rho$ and $v=l/t$.

\subsection*{Final network architecture}
The chosen DNN architecture is $N_l \times N_u = 6\times 32$.
It comprises $5569$ trainable parameters and occupies approximately $22$~kb of memory. This can be compared to the 4D 
reconstruction method~\cite{Lemasson2023a}, which utilizes matrices with dimensions $960 \times 450 \times 180 \times 260$, corresponding to  ($x_{f}$, $\theta_{f}$, $\phi_{i}$, $\theta_{i}$), after zero suppression 
and compression, occupying about $1$~Gb of memory. The speed of the 7DNN reconstruction method, encompassing both $\delta$ 
and $\varepsilon$ variables, is $1.4 \times 10^6$ events/s, making it suitable for efficient online and offline analysis.
Figure~\ref{fig:Conv} illustrates the training convergence of the $N_l \times N_u = 6\times 32$ deep neural network in terms of 
RMSD  for $\delta$ and $\varepsilon$.  RMSD$(\delta) = 0.5~\text{\textperthousand}$ and 
RMSD$(\varepsilon) = 0.16~\text{\textperthousand}$ are attained after $100$ iterations.

\begin{figure}[t]\center
\includegraphics[width=0.49\columnwidth]{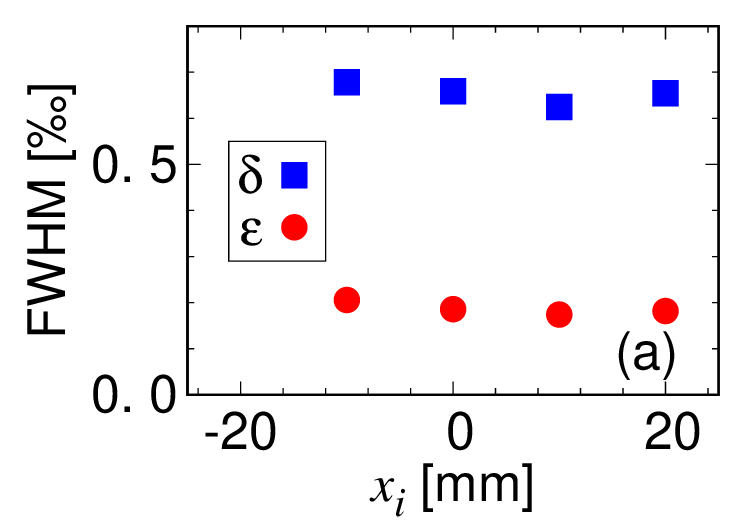}
\includegraphics[width=0.49\columnwidth]{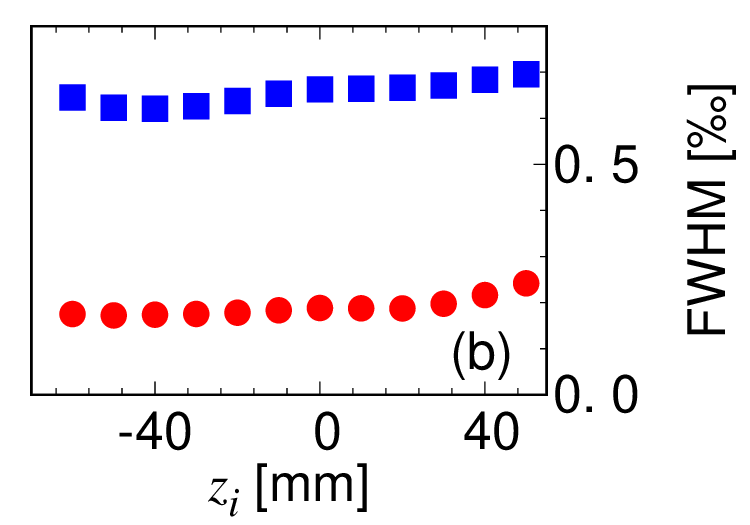}
\includegraphics[width=0.49\columnwidth]{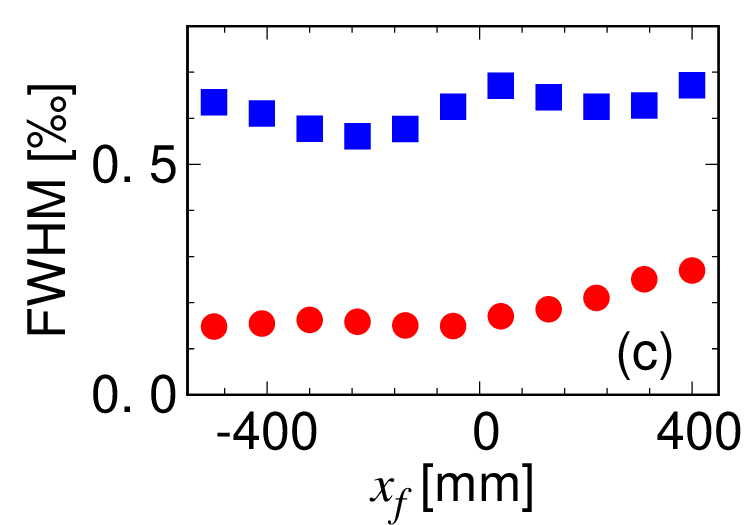}
\includegraphics[width=0.49\columnwidth]{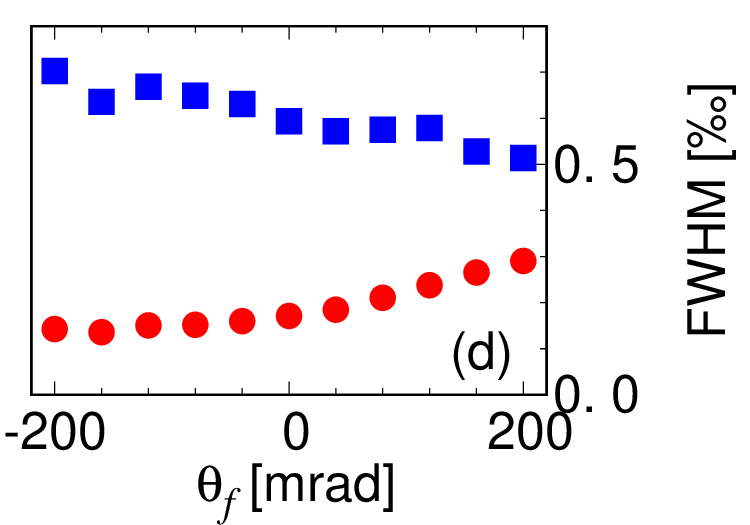}
\caption{Precision of the trajectory reconstruction: Full width at half maximum
FWHM  (blue squares) for the relative magnetic rigidity $(\delta)$
and  (red circles) for the relative trajectory length $(\varepsilon)$
of the distribution of the differences between the calculated by the ray-tracing code ($\delta_c$, $\varepsilon_c$) 
and reconstructed by 7DNN ($\delta_r$, $\varepsilon_r$)
as a function of (a) $x_{i}$, (b) $z_{i}$, (c) $x_{f}$ and (d) $\theta_{f}$, obtained using a complete dataset.
\label{fig:RecTst}}
\end{figure}

\section*{Results and performances}
\subsection*{Application of the model to simulated trajectories}
To assess the intrinsic precision of the 7DNN trajectory reconstruction method in relation to various key 
dimensions, the reconstruction method was applied to the complete set of trajectories generated by the ray-tracing 
code ZGOUBI. As mentioned above, the theoretical dataset included the relative rigidity in the range $\delta : [0.7,1.4]$.
The resulting relative trajectory length was in the range of $\varepsilon:[0.96:1.06]$. 
The overall values of FWHM$(\delta_c - \delta_r) = 0.65~\text{\textperthousand}$, equivalent to 
FWHM$(B\rho_c - B\rho_r) = 0.65$~mTm with $B_0=1$~Tm
and FWHM$(\varepsilon_c - \varepsilon_r) = 0.18~\text{\textperthousand}$,
FWHM$(l_c - l_r) = 1.4$~mm with $l_0=760$~cm,
were obtained for a complete dataset. The same values were obtained
considering only the part of the dataset relevant for the thin target.

In the following, the objective is to verify whether the obtained FWHM values for $\delta$ and $\varepsilon$, 
which are functions of the individual 
coordinates, are well represented by the overall result. The vertical coordinates $y_i$ and $\phi_i$ contribute minimally 
to $\delta$ and $\varepsilon$, and are therefore omitted. Since $\theta_i$ and $\theta_f$ exhibit a strong correlation, 
only $\theta_f$ will be analyzed for evaluating the performances of the DNN.
In Fig.~\ref{fig:RecTst} the resulting differences, obtained using a complete dataset,
between the calculated ($\delta_c$, $\varepsilon_c$), 
and reconstructed ($\delta_r$, $\varepsilon_r$) coordinates are analyzed in terms of the full width 
at  half maximum FWHM 
for $\delta_c - \delta_r$ (blue squares) and $\varepsilon_c - \varepsilon_r$ (red circles) 
as a function of different coordinates in the dispersive horizontal $xz$ plane.
The results are presented in Fig.~\ref{fig:RecTst} for (a) $x_{i}$, (b) $z_{i}$, (c) $x_{f}$, and (d) $\theta_{f}$ coordinates.
It is evident that the obtained FWHM are remarkably narrow, spanning from $0.5~\text{\textperthousand}$ to 
$0.7~\text{\textperthousand}$ for FWHM$(\delta_c - \delta_r)$ and from $0.1~\text{\textperthousand}$ to 
$0.3~\text{\textperthousand}$ for FWHM$(\varepsilon_c - \varepsilon_r)$. 
For $x_i$ and $z_i$, the FWHM closely follows the overall values obtained above. For positive $\theta_f$, 
the model achieves a better FWHM($\delta_c - \delta_r$) reaching $0.5~\text{\textperthousand}$. 
This can be correlated with the corresponding smallest optical aberrations (see Fig.~2 of Ref.~\cite{Lemasson2023a}), 
where $B\rho$ primarily depends on $x_f$ and less on $\theta_f$. The reason for a slight increase in 
FWHM$(\varepsilon_c - \varepsilon_r)\sim 0.3~\text{\textperthousand}$ for positive $\theta_f$ and positive $x_f$, 
which overlap in the low $B\rho$ region of the focal plane of VAMOS++, remains unclear but is nevertheless negligible.

In conclusion, the overall obtained result for the extended interaction volume is very satisfactory resulting
in very narrow widths.
The obtained FWHM values as a function of the individual parameters correspond well to the overall result.
Furthermore, the FWHM obtained for the extended interaction volume is equivalent to that obtained for the relevant 
portion of the dataset for the thin target.

\subsection*{Impact of the detector resolutions}

\begin{table}
\caption{Contribution of the detector resolutions: 
Result of the simulation, including for the parameters (column 1) 
their resolutions (column 2). 
The results are shown
in terms of FWHM for $\delta$, $\varepsilon$ and FWHM($A$)/$A$ for $A=80$ and $160$. 
In the first row the results are given without any detector resolutions. The subsequent rows only the resolution
of the corresponding parameter is included. The last column all resolutions are included.
Complete dataset was used.
}
\label{tab:tab1}
\center
\begin{tabular}{|l | l | l | l | l | l |}
\hline
 & \multicolumn{3}{|c|}{FWHM [\textperthousand]} & \multicolumn{2}{|c|}{FWHM/$A$ [\textperthousand]}\\
par. & par. & $\delta$ & $\varepsilon$ & $A=80$ & $A=160$\\
\hline
\multicolumn{2}{|c|}{without resolutions} & 0.65 & 0.18 & 1.10& 0.65\\
\hline
$x_i$ & 0.6 mm& 0.75 & 0.19 & 1.38 & 0.66\\
$y_i$ & 0.6 mm&  0.66 &  0.18 & 1.10 & 0.65 \\
$z_i$ & 1.1 mm& 0.66 & 0.24 & 1.13& 0.67\\
$\theta_i$  & 2.6 mrad& 0.82 & 0.34 & 2.10& 1.11\\
$\phi_i$  & 2.6 mrad& 0.66 & 0.18& 1.11 & 0.66\\
$x_f$ & 0.5 mm & 0.67 & 0.19 & 1.10& 0.66\\
$\theta_f$ & 1.1 mrad& 0.81 & 0.26 & 2.24 & 1.04 \\
$t$ & 1.0 ns& 0.65 & 0.18 & 5.29 & 3.81\\
 \hline
 \multicolumn{2}{|c|}{with all resolutions} & 1.32  & 0.42 & 6.32 & 4.17\\
 \hline
\end{tabular}
\end{table}

The theoretical dataset of trajectories was completed by the atomic charge state $q_{int}$ and 
the time-of-flight $t$, in accordance with the experimentally observed properties.
This enables the model to reconstruct also the atomic mass number 
$A$ using Eqs.~(\ref{eq:eq1}).
Table~\ref{tab:tab1} presents the results of the simulation, where the detector’s resolutions for each 
of the DNN input parameters and $t$ are utilized sequentially to study the network’s predictions.
To introduce the detector’s resolution for different parameters in the theoretical dataset obtained from ZGOUBI,
the values were randomly distributed over their respective resolutions, assuming a Gaussian distribution.
The experimental position and angular resolutions used and listed in the Table~\ref{tab:tab1} correspond 
to the typically obtained values for MWPC~\cite{Vandebrouck2016}. The time resolution correspond to the 
value obtained during in-beam experiments.
The first row corresponds to the simulation without any detector resolution, while the last row encompasses 
all resolutions combined. 
It is evident that among the network input parameters, the resolutions of the horizontal angles $\theta_i$ 
and $\theta_f$, along with the horizontal position $x_i$, have the most significant impact on the reconstructed 
FWHM. The dominant and limiting contributions arise from the time-of-flight, typically within the range $t:[180:260]$~ns,
which alone contributes more than a factor of 2 to the obtained FWHM($A$)/$A$ compared to any network input parameter. 
When all resolutions are combined, the FWHM($A$)/$80 = 6.32$~\text{\textperthousand} and FWHM($A$)/$160 = 4.17$~\text{\textperthousand}, which closely correspond to 
the experimental results presented below for the thin target experiment.

\subsection*{Application of the model to the experimental dataset}

\begin{figure}[t]\center
\includegraphics[width=1\columnwidth]{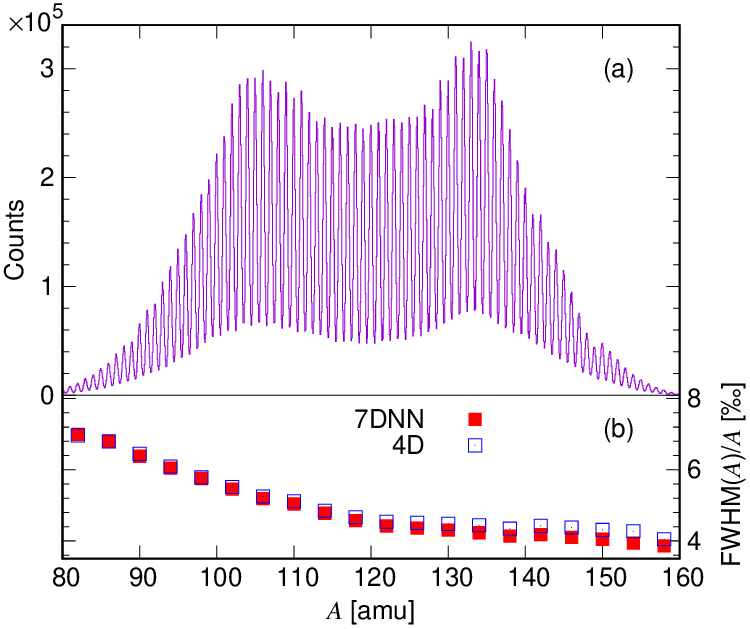}
\caption{
Application of 7DNN to experimental data:
(a) Atomic mass number spectrum obtained from the thin-target experiment (details provided in the text) using the 7DNN 
reconstruction method (left axis).
(b) Relative resolution of the atomic mass number FWHM$(A)/A$ (right axis) as a function of $A$ calculated using 7DNN (red filled squares) 
and 4D (blue open squares). 
\label{fig:SpectRes}}
\end{figure}

In Fig.~\ref{fig:SpectRes}(a), 
the experimental spectrum for the atomic mass number obtained using the 7DNN trajectory reconstruction is presented. 
The measured
($x_{i}$, $y_{i}$, $\theta_{i}$, $\phi_{i}$, $x_{f}$, $\theta_{f}$)  coordinates and $z_i = 0$ were used.
The resulting exceptional atomic mass number resolution is evident in the figure. The resolution can be quantified in terms 
of the full width at half maximum and is depicted as relative resolution FWHM$(A)/A$ in Fig.~\ref{fig:SpectRes}(b) by red filled squares as 
a function of $A$. The obtained relative resolution FWHM$(A)/A$ varied between $4~\text{\textperthousand}$ for the heaviest 
fission fragments and $7~\text{\textperthousand}$ for the lightest fission fragments. The difference in scale
between the widths for the intrinsic resolution shown in Fig.~\ref{fig:RecTst} 
(FWHM$(\delta_c - \delta_r) : [0.5, 0.7]~\text{\textperthousand}$, 
FWHM$(\varepsilon_c - \varepsilon_r): [0.1, 0.3]~\text{\textperthousand}$)
and shown in Fig.~\ref{fig:SpectRes}(b) 
FWHM$(A)/A : [4,7]~\text{\textperthousand}$ should be noted.
The experimentally obtained full width at half maximum is remarkably close to a constant FWHM$(A)= 0.6$~amu. The results obtained 
with the 7DNN method can be compared with those obtained for the same experimental dataset using the 4D method, 
as depicted by blue open squares in Fig.~\ref{fig:SpectRes}(b). It can be observed that for the light fission fragments, 
both methods yield comparable performances, while for the heavy fission fragments, the 7DNN method exhibits a slight 
advantage. Additionally, the results can be compared to those presented in Fig. 4 of Ref.~\cite{Lemasson2023a}. 
However, it is important to note that the experiment presented earlier was conducted in 2016, whereas the experiment 
presented here was performed in 2022~\cite{DataE826}, following several improvements to the VAMOS++ detection system.

\section*{Summary}

In summary, a novel trajectory reconstruction method for a large acceptance magnetic spectrometer VAMOS++ based on artificial deep neural 
networks was presented. The reconstruction method is based on seven-dimensional input from parameters required for the analysis of experiments 
involving reactions on extended target volumes. The networks were trained on the dataset of the trajectories generated by the ray-tracing code ZGOUBI. 
The most suitable architecture was determined based on their convergence and the reconstruction quality of the experimental data in terms of 
the atomic mass number. The architecture employed is $N_l \times N_u = 6\times 32$,  comprising $6$ layers of $32$ units (neurons) per layer, 
followed by a single output neuron. 
While the intrinsic reconstruction resolution for the relative magnetic rigidity, obtained through an application to the calculated trajectories, 
is (FWHM$(\delta_c - \delta_r) : [0.5, 0.7]~\text{\textperthousand}$, FWHM$(\varepsilon_c - \varepsilon_r): [0.1, 0.3]~\text{\textperthousand}$), 
the resolution resulting from an application to the experimental thin-target data is FWHM$(A)/A : [4,7]~\text{\textperthousand}$. 
The new 7DNN method demonstrates comparable or superior resolution to any previously employed trajectory reconstruction methods  for VAMOS++. 
The model exhibits remarkable speed, enabling the treatment of approximately $1.4\times 10^6$~events per second and minimal memory consumption. 

In the future, the reported method represents an opportunity for the reconstruction of trajectories, including extended beam spots.  This is particularly important for the forthcoming program that intends to utilize the voluminous gas target in conjunction with the  VAMOS++ magnetic spectrometer. It is anticipated that the optical axis  of the spectrometer and the beam axis will be rotated relative to one another. In this scenario, the assumption of the point-like beam interaction volume, upon which the preceding trajectory reconstruction methodologies were based, no longer holds. Consequently, it is imperative to explicitly incorporate the three-dimensional beam interaction volume, in addition to the previously utilized coordinates, into the newly developed trajectory reconstruction method. Furthermore, the method presented herein can also be employed for rapid simulation of the ion transmission within the VAMOS++ spectrometer. This presents novel opportunities for the development of a digital twin of the VAMOS++ spectrometer, which could be utilized for applications such as  determination of the absolute cross sections~\cite{Watanabe2013} 
or the measurements of the fission yields~\cite{Caamano2013}.

\section*{Data and software availability}
The dataset of trajectories utilized for training purposes is available at~\cite{Lemasson2025}. 
The training and reconstruction software are made available at~\cite{Rejmund2025}. 
The experimental data presented in Figure~\ref{fig:SpectRes} was acquired from the E826 GANIL experimental 
dataset, as referenced in~\cite{DataE826}.


\end{document}